\documentstyle{aipproc}

\begin{document}
\title{Estimates of Vertex Tagging Efficiencies at a Muon Collider Higgs
Factory}

\author{Bruce J. King}
\address{Brookhaven National Laboratory\\
email: bking@bnl.gov}\thanks{This work was performed under the auspices of
the U.S. Department of Energy under contract no. DE-AC02-76CH00016.}
\maketitle

\begin{abstract}
   Tagging efficiencies and purities are estimated for the decay modes
$\rm{H \rightarrow b \overline{b}}$,
$\rm{H \rightarrow \tau \overline{\tau}}$
and $\rm{H \rightarrow c \overline{c}}$ of Higgs bosons produced at
an s-channel muon collider Higgs factory.
\end{abstract}

\section*{Introduction}

  Measurement of the branching ratios (BR) of the Higgs boson would be an
important goal of a future s-channel muon collider Higgs factory.
This paper derives quick, heuristic estimates of the expected tagging
efficiencies and purities for the 3 main decay modes of a light,
Standard Model Higgs: 
$\rm{H \rightarrow b \overline{b}}$,
$\rm{H \rightarrow \tau \overline{\tau}}$
and $\rm{H \rightarrow c \overline{c}}$.
These estimates should
be useful as input to theoretical assessments of the physics capabilities
of Higgs factory muon colliders.

  The vertex tagging methods are similar for $\rm{b \overline{b}}$ events and
$\rm{c \overline{c}}$ events and these two modes are treated together in
the next section. The $\tau \overline{\tau}$ mode is then discussed in a
separate section, before ending with a short conclusion section.

\section{Charm and Beauty Tagging}

\subsection{Tagging Signatures}

  Higgs decays to c or b quark-antiquark pairs produce back-to-back 2-jet
events with a displaced vertex in each jet from the decay of the c or b
hadron. The goal of a vertex tagging algorithm is to find the displaced
vertices and also to distinguish the $\rm{b \overline{b}}$ events from
$\rm{c \overline{c}}$ events.
The figures of merit for the tagging algorithm are the overall tagging
efficiency for each of the 2 event types and the rejection factor against
the other, wrong event type.
In principle, one should also consider the rejection factor
for hadronic events that are neither $\rm{ b \overline{b}}$ nor
$\rm{c \overline{c}}$ but, in practice,
this is a much simpler task than distinguishing the 2 types with displaced
vertices.

  For b jets, the charged tracks in the displaced vertex will have
an invariant mass up to the mass of b hadrons -- about 5 GeV. There will
usually be several tracks.
B hadrons almost always decay into a charm hadron plus additional
hadrons, so some of the charged tracks will likely form a tertiary
vertex downstream from the B decay vertex. If a ``topological vertexing''
algorithm such as ZVTOP~\cite{zvtop} is used then this tertiary vertex
may be reconstructed. The characteristic displacement length of the vertex
is $\beta  \gamma . \rm{c} \tau$, where $\rm{c} \tau$
is 450 microns. $\beta \gamma$ would typically be about 7 for a 100 GeV Higgs
decay at rest -- corresponding to about 70\% of the quark energy or
35\% of the Higgs mass, as is
seen in Z decays to b quarks -- so the characteristic displacement length
is about 3 mm.

  The displaced vertices from charm typically have a lower multiplicity
and, more significantly, a lower invariant mass than B decays since the
mass of all weakly decaying charm hadrons is less than 2 GeV. In analogy
to Z decays, the charm hadrons should typically carry about 50\% of
the c quark energy. Combining this with $\rm{c} \tau = 317$ (124) microns
for charged (neutral) D mesons gives characteristic decay lengths of
4 mm (1.7 mm). The lower multiplicity and lower invariant mass of
charm decays makes them more difficult to tag than B decays, even for
the longer lived charged D mesons.

\subsection{Jet Tagging Efficiencies}

  An estimate for jet tagging efficiencies and purities was made based
on studies done for the proposed DESY 500 GeV center-of-mass electron-positron
linear and presented in the Conceptual Design Report (CDR)~\cite{desycdr}.
This study was performed for the following scenario:
\begin{enumerate}
  \item  a 3 layer barrel vertex detector with its innermost layer a
cylinder at either 1.0 cm or 2.2 cm from the interaction point (ip)
  \item pixel tracking elements. Studies were performed for both
charge coupled devices (CCD's) and active pixel sensors (APS's)
  \item a 50--50 mix of b and c jets
  \item vertex reconstruction using the SLD topological vertexing
algorithm ZVTOP~\cite{zvtop}
  \item the central region of the detector was considered.
\end{enumerate}

  A muon collider Higgs factory will have worse backgrounds than a linear
electron-positron collider and two adjustments were made to allow for this.
Firstly, APS's were assumed to be the tracking technology because CCD's
are presumably too susceptible to radiation damage for use in the muon
collider tracking environment and, secondly,
the radius of the innermost tracking layer was assumed to be 5 cm,
consistent with background studies for muon colliders~\cite{iuliu}. This
requires an extrapolation of the DESY studies.

  With these adjustments, it is assumed that the Higgs factory jet tagging
performance can be estimated using the DESY studies. This implicitly assumes
that 2 further differences are not important in the study. Firstly, it is
assumed that the effect of the higher uncorrelated ``fake hit'' density
in a muon collider can be minimized in a well designed vertex detector using
the redundancy afforded by the several vertexing layers. Secondly, the
Higgs factory muon collider will have a beam spot size in the hundreds of
microns which, in contrast to the smaller spot at a linear collider, will give
little useful vertex
constraint to assist the vertexing algorithm. The lack of a vertex constraint
is important today's collider detectors  that use 2-D vertexing with silicon
microstrips and only 2 or 3 layers of vertexing e.g. the LEP and Tevatron
detectors. For these geometries the additional constraint would be very
helpful. However, it is probably reasonable to assume that the loss would be
much less in a multi-layer 3-D pixel vertex detector such as one would expect
in a future muon collider.

  The purity vs. efficiency curves from the DESY study are shown in
figure 2.2.2 of the DESY CDR~\cite{desycdr}. All of the curves have a fairly
flat purity out to a certain efficiency then dive quite steeply,
so little is lost in either purity or efficiency by choosing
the efficiency and purity values at this lip of the curve. These values
for 1.0 and 2.2 cm inner radii and b-- and c--tagging are given in
table~\ref{tab:effpur}. The table also gives the assumed extrapolation to
the 5.0 cm radius at a muon collider.

  The extrapolated values were estimated
by noting that the extrapolation ratio 2.2:5.0 cm is similar to the
ratio between the 2 DESY values, 1.0:2.2 cm. The extrapolation to the
muon collider scenario is rather modest since the observed changes between
1.0 cm and 2.2 cm are relatively small. Thus, it is considered that this
specific choice of extrapolation introduces little uncertainty and
should be entirely adequate for estimates of the physics potential of the
Higgs factory muon collider -- certainly until a detailed detector design
and simulation is arrived at.

\begin{table}[ht!]
\caption{Jet tagging efficiencies and purities for an equal sample of
b and c jets in the central region of the vertex detector. The values
for 1.0 cm and 2.2 cm were read off from the curves for APS's in figure
2.2.2 of the DESY CDR and these values have been extrapolated to give
efficiencies and purities at 5.0 cm.}
\begin{tabular}{|c|ccc|}
\hline
tag   & radius &  efficiency (e)  & purity (p)  \\
\hline
\hline
b-tag & 1.0 cm &  63 \%  & 97 \% \\
b-tag & 2.2 cm &  59 \%  & 97 \% \\
b-tag & 5.0 cm &  55 \%  & 97 \% \\
\hline
c-tag & 1.0 cm &  46 \%  & 74 \% \\ 
c-tag & 2.2 cm &  42 \%  & 71 \% \\
c-tag & 5.0 cm &  38 \%  & 68 \%
\label{tab:effpur}
\end{tabular}
\end{table}

The purities in table~\ref{tab:effpur} refer
to an equal mixture of c and b jets.
For dealing with an arbitrary mixture of b and c jets it is useful to
convert these purities into efficiencies for tagging the wrong type
of jet. For an equal sample of b's and c's and subscript $i$ ($j$)
denoting the intended (wrong) tag it is clear that:
\begin{equation}
p_i = \frac{e_i}{e_i + e_j},
\end{equation}
so solving for $e_j$ gives:
\begin{equation}
e_j = e_i \times (\frac{1}{p_i} - 1)
\end{equation}

 This gives the jet tagging efficiencies of table~\ref{tab:jeteff} for
the central region of the detector.

\begin{table}[ht!]
\caption{Jet tagging efficiencies for b-- and c--tagging in the central
region of the vertex detector for a Higgs factory muon collider. The
innermost tracking layer of the vertex detector is assumed to be at 5.0 cm
from the ip.}
\begin{tabular}{|c|cc|}
\hline
tag   & eff. for b jets &  eff. for c jets \\
\hline
b-tag  &  55 \%  & 2 \% \\
c-tag  &  18 \%  & 38 \%
\label{tab:jeteff}
\end{tabular}
\end{table}

\subsection{Event Tagging Efficiencies}

  The jet tagging efficiency must be converted to an event tagging efficiency.
Each Higgs decay to $\rm{ b \overline{b}}$ or $\rm{c \overline{c}}$ will
produce an event with two
approximately back-to-back jets with the same quark flavor. In this subsection,
the event tagging efficiency and rejection factor against wrong flavor events
is calculated from the assumed jet tagging efficiencies using a simple
algorithm for combining the jet tagging information from the two jets. Several
simplifying approximations are used.

  Since the jets are fairly back-to-back it will be assumed that either both
jets are central or neither jet is. Therefore, the calculation is done
assuming central events and then the efficiency is multiplied by an overall
geometrical acceptance factor of $\cos (\pi /4) = 1/\sqrt{2}$. Since the
Higgs decay is isotropic this is roughly equivalent to assuming that jets
more than 45 degrees from the beam direction can be considered central enough
for reliable vertex tagging.

\begin{table}[ht!]
\caption{Probabilities for tagging 0,1 or 2 jets if the jet tagging
efficiency is $e$.}
\begin{tabular}{|c|c|}
\hline
number of tagged jets   & probability \\
\hline
0  & $(1-e)^2$ \\
1  & $2e(1-e)$ \\
2  & $e^2$
\label{tab:prob}
\end{tabular}
\end{table}

  The simple event tagging strategy is to require either one or both of
the 2 jets to be tagged correctly and also to require that neither was
tagged as the incorrect flavor. From table~\ref{tab:prob}, it is seen that the
probability for the first condition is $1-(1-e_i)^2$ and for the second
condition is $(1-e_j)^2$. (As before, $i$ denotes the correct jet tag
and $j$ the incorrect tag.) If the simplifying assumption is made that
the two probabilities are independent and the $1/\sqrt{2}$ geometrical
acceptance is included then the efficiency, $E_{ii}$, for correctly
tagging the event is:
\begin{equation}
E_{ii} = \frac{1}{2} \times [1-(1-e_i)^2] \times (1-e_j)^2.
\end{equation}

  Clearly, the probability, $E_{jj}$ for tagging the wrong event type is
obtained by simply swapping the $i$'s and $j$'s:
\begin{equation}
E_{jj} = \frac{1}{2} \times [1-(1-e_j)^2] \times (1-e_i)^2,
\end{equation}
and the rejection factor $R$ against the wrong flavor event is
defined naturally to be
\begin{equation}
R = \frac{E_{ii}}{E_{jj}}.
\end{equation}

  The $\rm{ b \overline{b}}$ and $\rm{c \overline{c}}$ tagging efficiencies
and rejection factor against the wrong flavor are given in
table~\ref{tab:effrej}.

\begin{table}[ht!]
\caption{Event tagging efficiencies and wrong-flavor rejection factors
for $\rm{ b \overline{b}}$ and $\rm{c \overline{c}}$ events at a
Higgs factory muon collider. A geometrical
acceptance factor of $1/\sqrt{2}$ is included in the efficiencies.}
\begin{tabular}{|c|cc|}
\hline
event type  &  efficiency, $E_{ii}$   &  rejection factor, $R$ \\
\hline
$\rm{ b \overline{b}}$  &  54 \%  & 50 \\
$\rm{c \overline{c}}$  &  42 \%  & 8
\label{tab:effrej}
\end{tabular}
\end{table}

\section{Tau event tagging}

  The tagging of $h \rightarrow \tau \overline{\tau}$ events appears
to be much easier than the tagging of quark jets. An s-channel
Higgs factory will produce Higgs at rest and not in association with other
particles, so the geometry should be identical to the
$Z \rightarrow \tau \overline{\tau}$ events seen in the LEP and SLD detectors
operating at the Z pole energy. These distinctive events consist of almost
back-to-back high energy tracks emanating from the ip. Each side is usually
a single track -- i.e. a ``1-prong'', with an 85.5 \% probability per side
-- with almost
all of the remainder being tightly collimated ``3-prong'' jets. The tau
lifetime is long enough ($c \tau = 88$ microns) that the slight offset
of the prongs from the ip should be observable in a precise vertex detector,
but vertexing information should not generally be required to identify
this event sample. Therefore, the purity of the sample should be
close to 100 \% and the efficiency should be dominated by the geometrical
acceptance of the central tracker. For all practical purposes, physics
studies could reasonably assume a purity of 100\% and a conservative
efficiency of $\cos (\pi /4) = 0.71$.

\section{Conclusions}

  Heuristic estimates have been made of tagging efficiencies and purities
in an s-channel muon collider Higgs factory for the 3 main decay modes of
a light Higgs boson. The estimates for the $\rm{ b \overline{b}}$ and
$\rm{c \overline{c}}$ modes are given in table~\ref{tab:effrej}, while
the $\tau \overline{\tau}$ mode is assumed to have approximately a
71 percent efficiency with essentially 100 percent purity.

\end{document}